# Atomically localized ingredient-dependent interface phonon in heterogeneous solids

Mei Wu[1,2,#], Ruochen Shi[1,2,#], Ruishi Qi[1,2,3,#], Yuehui Li[1,2], Tao Feng[4,5], Bingyao Liu[2], Jingyuan Yan[2], Xiaomei Li[1,2], Zhetong Liu[2], Tao Wang[2], Tongbo Wei[4,5], Zhiqiang Liu[4,5], Jinlong Du[2], Ji Chen[6,7], & Peng Gao[1,2,7,8*]

**Phonons are the primary heat carriers in non-metallic solids. In compositionally heterogeneous materials, the thermal properties are believed to be mainly governed by the disrupted phonon transport due to mass disorder and strain fluctuations, while the effects of compositional fluctuation induced local phonon states are usually ignored. Here, by scanning transmission electron microscopy electron energy loss spectroscopy and sophisticated calculations, we identify the vibrational properties of ingredient-dependent interface phonon modes in $Al_xGa_{1-x}N$ and quantify their various contributions to the local interface thermal conductance. We demonstrate that atomic-scale compositional fluctuation has significant influence on the vibrational thermodynamic properties, highly affecting the mode ratio and vibrational amplitude of interface phonon modes and subsequently redistributing their modal contribution to the ITC. Our work provides fundamental insights into understanding of local phonon-boundary interactions in nanoscale inhomogeneities, which reveal new opportunities for optimization of thermal properties via engineering ingredient distribution.**

[1] International Center for Quantum Materials, Peking University, Beijing 100871, China.
[2] Electron Microscopy Laboratory, School of Physics, Peking University, Beijing 100871, China.
[3] Physics Department, University of California at Berkeley, Berkeley 94720, USA.
[4] Research and Development Center for Semiconductor Lighting Technology, Institute of Semiconductors, Chinese Academy of Sciences, Beijing 100083, China
[5] Center of Materials Science and Optoelectronics Engineering, University of Chinese Academy of Sciences, Beijing 100049, China
[6] State Key Laboratory for Mesoscopic Physics, School of Physics, Peking University, Beijing 100871, China
[7] Collaborative Innovation Center of Quantum Matter, Beijing 100871, China.
[8] Interdisciplinary Institute of Light-Element Quantum Materials and Research Center for Light-Element Advanced Materials, Peking University, Beijing 100871, China.
# These authors contribute equally to this work.
* Corresponding author. E-mail: p-gao@pku.edu.cn.

Thermal conductivity describes the heat dissipation capability inside materials, being a key factor in thermoelectric energy conversion [1,2] and thermal management of electronic devices [3,4]. For heterogeneous materials, the fluctuations in local concentration typically hold the key to determine thermal properties [1,5,6] even and particularly at the nanoscale. For example, a great many recent advances in designing high-performance thermoelectric materials are linked to the modulation of nanoscale ingredient inhomogeneity [5,7]. These nanoscale control such as doping, alloying and nanoprecipitation will disrupt the phonon transport and subsequently decrease the thermal conductivity [8,9]. Previous conventional consideration about these nanoscale phenomena mainly focuses on the mass disorder and strain fluctuations which assumes the random mixture of atoms with different masses and volumes in a lattice [7,10] and has been well-studied [11,12]. However, the influence of boundaries/interfaces induced by ingredient variation in inhomogeneous solids is usually oversimplified [13-15], much less a clear cognition of the relation between the ingredient changes and interface thermal conductance (ITC) changes. In fact, the previous theoretical work observed the superlattice thermal conductivity was lower than that of the alloy with the same mass ratios [16,17], indicating the important contributions of the ITC and thus calling for unraveling the unique thermal properties of interfaces.

The proper knowledge of the thermal transport mechanisms occurring at the compositional interface requires considering the local phonon modes. Previous extensively accepted descriptions about the ITC, such as the acoustic mismatch model and the diffuse mismatch model for the phonon transport, mainly consider the mismatch of the phonon density of states (DOS) between constituent bulk materials [18]. In fact, the thermal properties of materials are rooted in their microstructures. Due to different bonds at the interface, unique eigen solutions are required to describe the equations of atomic vibrational motion compared to the bulks [19]. Indeed, there are more and more theoretical work concerning the important role of such localized modes (eigenmodes with large displacements for interfacial atoms) on the ITC [20-22]. Unfortunately, experimentally the heat dissipation at an interface still cannot be dealt with confidently, mainly owing to the lack of ability in probing local phonon DOS, dispersion relation and scattering mechanism at the interface, arising from the inadequate spatial resolutions in conventional experimental measurements [23-25]. How to bridge the theoretical calculation with experimental results is also essential for well understanding the nature of phonon-interface scattering.

Here, we use the advanced vibrational electron energy loss spectroscopy (EELS) technique in scanning transmission electron microscopy (STEM) [26-33] combined with sophisticated density functional perturbation theory (DFPT) calculations and molecular dynamics (MD) calculations, trying to provide a direct physical picture of the correlation among the ingredient change, lattice dynamics and thermal properties at the atomic scale. We choose the $Al_xGa_{1-x}N$-based interface system, which is extensively used in highly integrated microelectronic products and optical devices [34,35]. To reveal the fundamental mechanism of phonon-interface scattering, we first study the GaN/AlN interface, confirming the existence of localized interface phonon modes from the atomically resolved phonon DOS measurements and mapping the phonon dispersion across the interface with nanometer atomic resolution. These interface phonon modes are highly localized within ~ 1.5 nm and make significant contribution to the ITC. For the $Al_xGa_{1-x}N$/AlN system, it shows similar interface phonon characteristics with GaN/AlN interface. However, these interface phonon modes are sensitive to compositional fluctuation. We demonstrate that as the Al fraction increases, the number of states and vibrational amplitude for the localized modes are reduced, subsequently showing lower degree of interaction with other modes and contributing less ITC. Our work provides insights of the fundamental phonon transport physics with the nanoscale and atomic-scale compositional fluctuations in heterogeneous materials. The engineering of vibrational thermodynamics over compositional intertwining can also help to optimize efficient thermal managements in material science.

**Interface lattice dynamics across the nitride interface**

Figure 1 gives a schematic illustration of the distinctive nanostructured geometry in heterogeneous media. Ideally, the different ingredients are fully mixed (Fig. 1b), introducing mass contrast and strain field fluctuations. In practice, the local concentrations are always varied in stable compounds. For example, nanophase segregations and nanoprecipitations commonly exist in thermoelectric materials, ubiquitously resulting in the emergence of interfaces. Generally, we can separate the inhomogeneity with different kinds of boundaries, some of which shows fully sharp ingredient-separated interface structure (Fig.1c) and others shows element interdiffusion (Fig. 1d). As observed by previous theoretical work, the superlattice thermal conductivity was lower than that of the alloy with the same mass ratios [16,17], indicating the compositional interface plays an important role in the thermal conductivity.

Hence, in order to give a thorough description of the phonon-interface scattering in nanoscale compositional fluctuations, we choose the $Al_xGa_{1-x}N$-based interface as a model system and first study the most essential GaN/AlN interface. The atomically resolved high angle annular dark field (HAADF) image (Supplementary Fig. 1) and the integrated differential phase contrast (iDPC) image (Fig. 2a) show the atomically sharp and coherent interface. To obtain GaN/AlN interface vibration characteristics, the phonon spectra across the interface (Fig. 2b) are recorded in the off-axis setup with a large convergence angle, under which the acquired signal is close to local phonon DOS [27,33,36]. The experimental details are included in the Methods and Supplementary Fig. 1. For better describing the phonon DOS variation, we extract the peak energy (Fig. 2c) by least-square fitting the measured spectrum to a sum of Lorentzian peaks. In the GaN side, the spectrum displays four major peaks at around 24 meV, 39 meV, 76 meV and 85 meV, while in the AlN side, three peaks at around 38 meV, 65 meV and 84 meV are distinguishable. When approaching the interface, the spectrum features change.

We then performed DFPT calculations, aiming at giving accurate and detailed phonon scattering mechanisms. As shown in Supplementary Fig. 2a, bulk GaN and AlN have similar phonon dispersion profiles with large degree of overlap of phonon DOS, except that GaN features an obvious optical-optical gap due to a larger atomic mass difference between cation and anion [37]. The corresponding calculated projected DOS of each atom layer across the interface is illustrated in Fig. 2d, which exhibits good agreement with the experimental data (Fig. 2b). Specifically, for GaN adjacent to the interface, the acoustic branch (at ~24 meV) has slight blue shift. Moreover, as marked by the black arrows in Figs. 2b-d, the GaN layer adjacent to the interface vibrate at frequency above its maximum vibrational frequencies ($\omega_{max,GaN}$) (from 85 meV to 89 meV approaching the interface), which is similar to the localized phonon modes at the Si/Ge interface with the frequency between the $\omega_{max,Si}$ and $\omega_{max,Ge}$ [38]. Besides, the projected DOS at 80 meV is prominent in the GaN layer adjacent to the interface, accounting for the experimental observation on the blueshift of vibrational energy loss peak at 76 meV near the interface. On the AlN side, a slight red shift is observed for the peak at 65 meV approaching the interface. Such enhancement of phonon DOS overlap at the interface may mitigate the phonon mismatch between two materials and thus improve phonon transmission at the interface.

For better understanding the local phonon behaviors at the interface, we extract the measured spectra (Fig. 2e) and calculated projected DOS (Fig. 2f) with the beam located in GaN (blue curve), in AlN (red curve), and at the interface (purple curve). We fit the interface spectrum with the linear combination of two bulk spectra to match the acquired spectra best. Then the residual part (black line) means the exceptional properties that only exist at the interface, where the experimental and calculated results show decent agreement. The corresponding experimental fitting residuals are also shown in Supplementary Fig. 2b and 2c. As can be seen, despite the large overlap of phonon DOS among GaN and AlN, the heterointerface between them certainly shows unique phonon modes. In order to further validate their presence, we extracted the energy-filtered EELS maps shown in Fig. 2g and 2h. The intensity maps for 30-34 meV energy window (left panel in Fig. 2g) show an enhanced intensity at the interface within ~1.5 nm in space. Such enhancement is also consistent with the positive residual and the higher intensity of interface spectrum at 30-34 meV energy window compared with bulk sides (marked by the black arrow in Figs. 2e and 2f), corresponding to the localized modes with enhanced vibration at the interface. What's more, we observe decreased intensity at the interface at 68-72 meV (left panel in Fig. 2h), corresponding to the negative residual and implying the isolated modes that both sides vibrate but the vibration amplitude is reduced at the interface. Their presence is also confirmed by the corresponding calculations of phonon eigenvectors (right panel in Figs. 2g and 2h). Besides these two phonon modes, other classes of interface phonon modes (extended modes and partially extended modes, representing vibrations spread over the whole system and vibrations mainly exhibit at one side, respectively) maps are also recognized (Supplementary Fig. 3). Due to the limited energy resolution and large overlap of phonon DOS among the bulks and interface, we cannot precisely extract the spatial distribution of each localized mode and isolated mode experimentally but we extract the calculated phonon eigenvectors for the corresponding region of positive and negative residual (Supplementary Fig. 4), showing consistent vibration amplitude distribution.

Furthermore, we explore the dependence of vibrational energy on momentum across the interface with nanometer spatial resolution using the four-dimensional EELS (4D-EELS) technique [32,39]. Supplementary Fig. 5 shows typical phonon dispersion curves along ΓKMKΓ viewing from [1$\bar{2}$0] direction from experiments and corresponding simulations for GaN, in AlN, and the interface respectively. The reasonable agreement between the experiment and simulation further verifies the presence of complicated vibrational features at the interface which are different than those in the bulks.

**Ingredient-dependent phonon-boundary phonon in inhomogeneous nitride**

The following crucial question is the influence of compositional fluctuation and whether we can engineer the interface phonon behavior. Therefore, we explore the $Al_xGa_{1-x}N$/AlN system where one side ingredient is varied. Fig. 3a displays the atomic-resolution HAADF image, showing identical interface structure with the GaN/AlN system. The corresponding elemental distribution is also revealed by the line profiles of energy dispersive X-ray spectroscopy (EDS) (Fig. 3b). The off-axis phonon spectra across the interface (Fig. 3c) show similar tendency with the GaN/AlN system but with unobtrusive fitting residual as shown in Fig. 3d. To systematically and qualitatively interpret the vibrational spectra of $Al_xGa_{1-x}N$/AlN system, we built supercell based on the above DFT-optimized atomic configuration and randomly replaced Ga atoms by Al atoms with ratio x in each atomic plane. The corresponding typical model are illustrated in Fig. 3b and the calculated details are summarized in the Methods. By calculating the lattice dynamics of compositional-change $Al_xGa_{1-x}N$/AlN system (x=0, 0.2, 0.5, 0.8), we first extracted the mode numbers of different interface phonon modes (Fig. 3e). As the Al fraction (x) in $Al_xGa_{1-x}N$ increases, the mode ratios of four types of interface phonon modes are varied where the number of states for localized modes is decreased and extended modes gradually dominate the interface phonon transition. The interface vibration amplitude of different modes follows a more concentrated distribution around the average amplitude (Fig. 3f) with increasing x, indicating the reduced amplitude of localized modes. These decreased number of states and vibration amplitude for localized modes may account for the diminished fitting residual of interface particular ingredient for the $Al_xGa_{1-x}N$/AlN system.

**Influence of interface vibration on interface heat conduction**

In order to clarify the influence of such interface phonon modes on the ITC, we performed molecular dynamics (MD) simulations using the interface conductance modal analysis (ICMA) formalism (Fig. 4) [40]. Firstly, we calculated the ITC decomposed into correlation integral map between phonon modes sorted by their interfacial amplitude (Fig. 4a and Supplementary Fig. 6). The eigenvectors are multiplied by a global scaling factor, such that the mean value of all modes' interfacial amplitude is equal to 1, labeled as the black dashed line in Fig. 4a and Supplementary Fig. 6. The greater value of normalized amplitude indicates the stronger vibration at the interface. Fig. 4a shows the correlation integral map of the GaN/AlN system. The higher intensity at the region of localized modes (normalized amplitude >1, right top region) illustrates those interfacial vibrations actually enhance the contributions to the thermal conductivity. While the isolated modes (at left bottom of the map) show few contributions to the ITC. For the $Al_xGa_{1-x}N$/AlN system (Supplementary Fig. 6), as the Al fraction increases, we observed the lower intensity of the right top region in correlation integral maps, indicating the decreased interaction between localized modes and other modes. Then we extract modal contributions to ITC associated with the four classes of interface phonon modes as shown in Figs. 4b and 4c. Due to the large overlap of phonon DOS among $Al_xGa_{1-x}N$ based compounds, the extended mode mainly contributes to the total ITC. Moreover, the contribution of extended modes is monotonically higher while the influence of other three modes decrease as the Al fraction increases. Interestingly, the per-mode contribution of extended modes (Fig. 4c) is almost invariable with the compositional change. For localized modes, they own the highest contribution to ITC on per mode basis (~ 60.8%) at the GaN/AlN interface, similar to previous report on other interface systems [32,38,41-43]. As the Al fraction increases, their per-mode contribution to ITC decrease, consistent with the reduced intensity of correlation integral map with the Al fraction increases. By now, we establish the direct picture of the compositional fluctuation contributing to thermal transports across the interface. It can be seen that even a subtle concentration variation will highly affect the mode ratio and thermal conductivity contribution of different interface phonon modes, affecting the ITC.

**Conclusion**

In conclusion, by systematically investigating the phonon-interface scattering in the compositional varied $Al_xGa_{1-x}N$/AlN system, we reveal the effects of local composition heterogeneity induced interfaces on the phonon transport and advance the understanding of the local thermal conductivity in practical nonuniform solids. With the Al fraction increases, the smaller mismatch between the two adjoined bulks results in the reduced vibrational mode numbers and amplitude of those localized modes at the interface, subsequently contributing smaller to ITC. Besides to the well-established knowledge of effects of mass disorder and strain fluctuations on local thermal properties, this work suggests that the compositional interface also play an important role in the local thermal conductivity. The tunable interfacial thermal characteristics via compositional fluctuations tailoring interface phonon modes also provide new insights into the engineering of thermal properties in heat dissipation and energy conversion materials and devices.

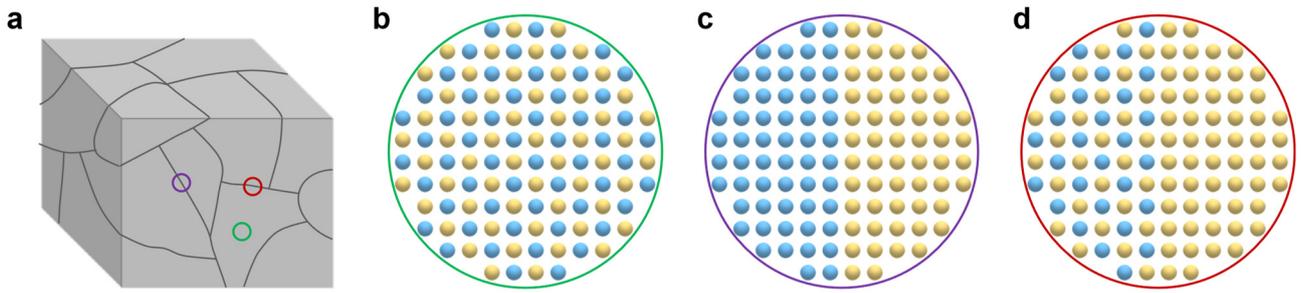

**Fig.1 Schematics of typical regions in heterogeneous medias. a**, Schematics of heterogeneous medias with various boundaries/interfaces. **b**, Schematics of fully disordered heterogeneous medias. **c**, Schematics of sharp compositional interface. **d**, Schematics of interface with chemical mixing in one side.

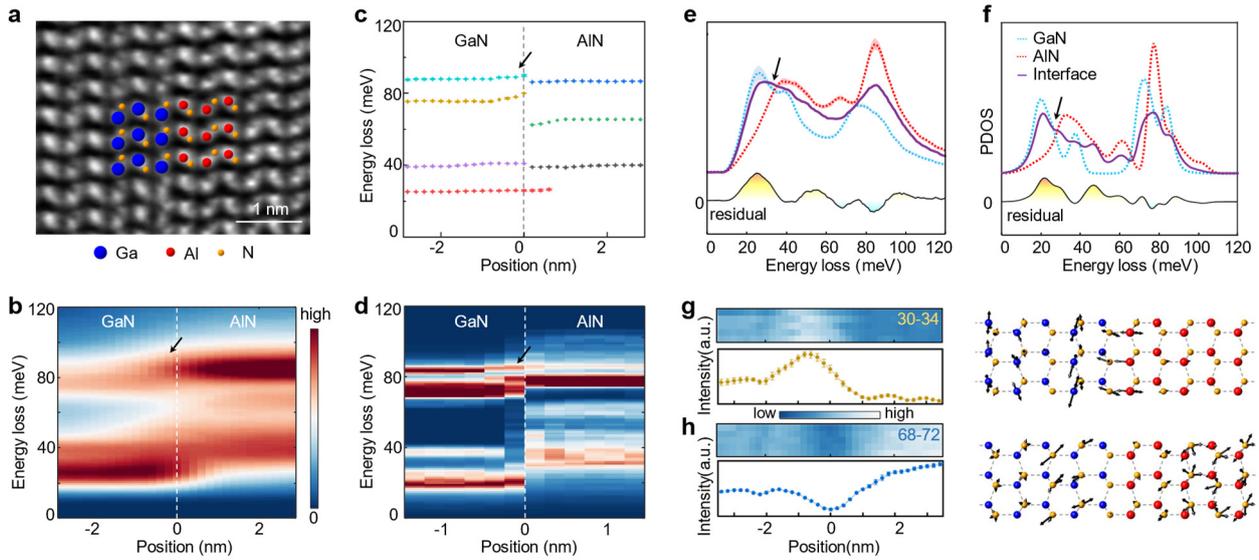

**Fig. 2 Interface vibration at the GaN/AlN heterojunction. a**, iDPC image of the GaN/AlN interface viewed along [110] zone axis, illustrating the atomic structure of interface. Blue, red and yellow balls represent Ga, Al and N atoms respectively. **b**, Spectrum profile of measured off-axis phonon spectra across the interface. **c**, Fitted peak positions of phonon spectra in Fig. 2b by least square fitting methods. **d**, Calculated phonon DOS projected on atom layers across the interface. **e,f**, Extracted EEL spectra (e) and calculated projected DOS (f) from GaN (blue), AlN (red), and the interface (purple). The black curve indicates the interface component that cannot be expressed as a linear combination of two bulk spectra, showing similar trends. **g,h**, Spatial distribution mapping and corresponding intensity line profiles of localized (g) and isolated modes (h) extracted from Fig. 2b. The energy integration windows are labeled in meV. The corresponding calculated phonon eigenvectors are shown in the right panel respectively. Arrows illustrate the vibration amplitude of each atom.

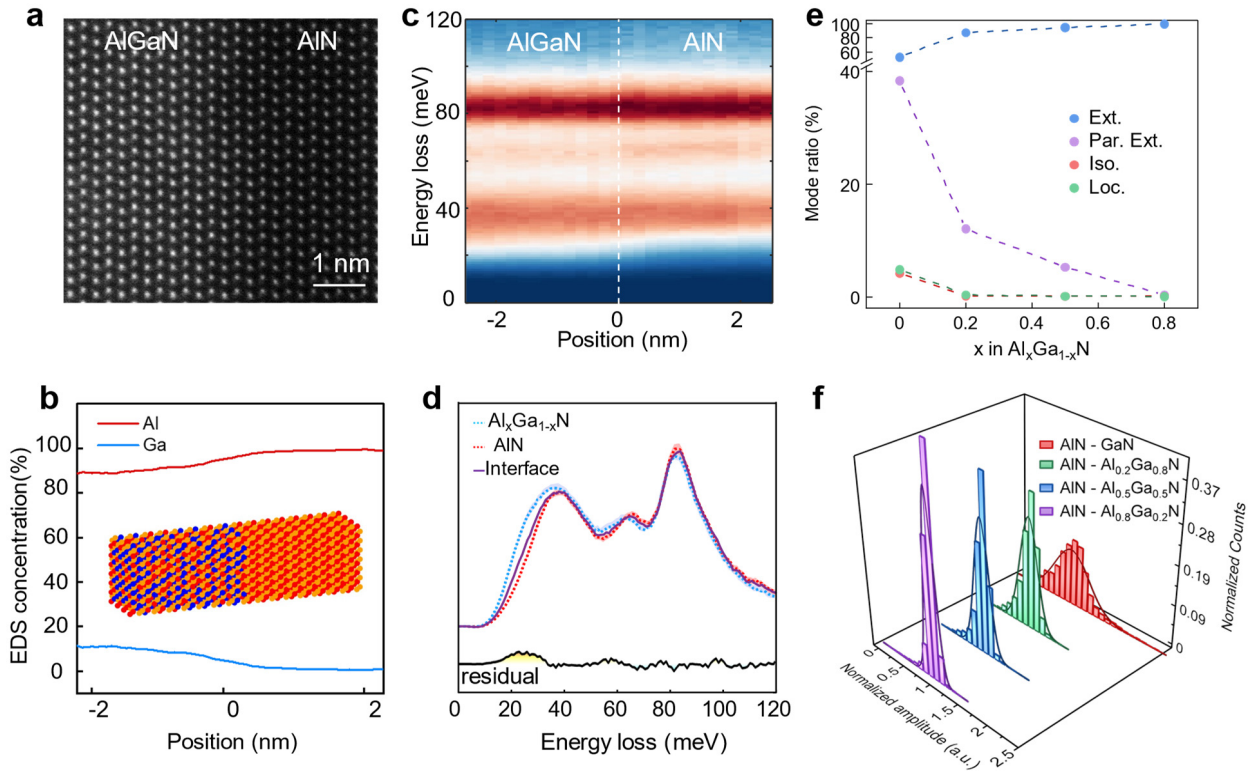

**Fig. 3 Interface phonon mapping for the Al$_x$Ga$_{1-x}$N/AlN interface. a,b** HAADF image (a) and corresponding EDS mapping (b) of the Al$_x$Ga$_{1-x}$N/AlN interface. **c**, Line profile of measured off-axis phonon spectra across the interface. **d**, Extracted EEL spectra from Al$_x$Ga$_{1-x}$N (blue), AlN (red), and the interface (purple). The black curve indicates the interface component that cannot be expressed as a linear combination of two bulk spectra. **e,** Calculated number of states for the four different classes of interface phonon modes with the change of Al fraction in Al$_x$Ga$_{1-x}$N. Ext: Extended modes. Par. Ext.: Partial Extended modes. Iso: Isolated modes. Loc: Localized modes. **f**, Calculated distribution of normalized interface vibration amplitude with the change of Al fraction in Al$_x$Ga$_{1-x}$N.

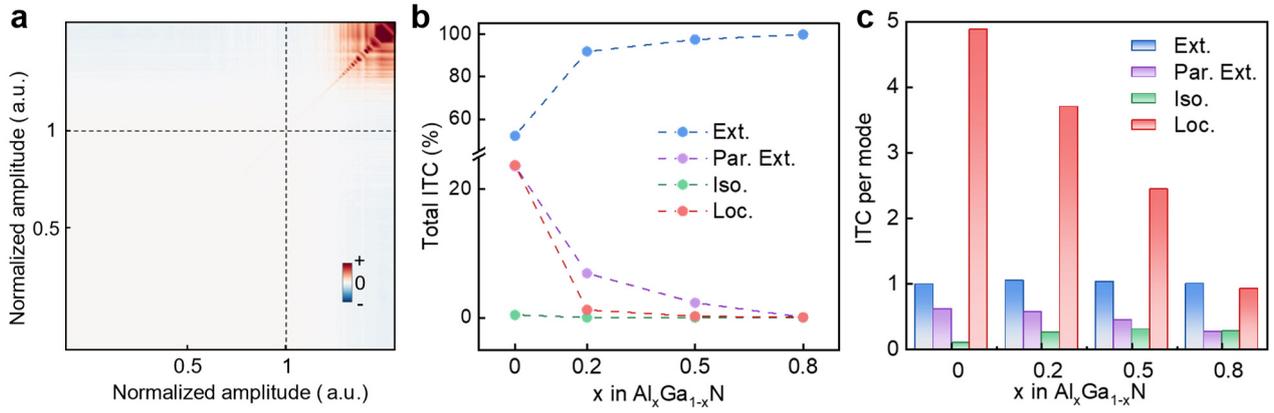

**Fig. 4 Decomposition of interface thermal conductance into modal heat flux correlation integrals.**
**a**, The magnitudes of the per-mode contribution to ITC sorted by their amplitude at the GaN/AlN interface. The axis illustrates the scale of interfacial amplitude. The dashed black curves represent the mean value of bulk vibration. A value greater than 1 means an enhanced vibration at the interface and a value smaller than 1 indicates a reduced vibration at the interface. The enhanced intensity in the top right-hand corner of the mapping signifies that large amplitude of vibrations at the interface greatly contributes to ITC, while modes with reduced amplitude at the interface show few contributions to ITC. **b**, The changes of contribution to total ITC from four different classes of interface phonon modes with the Al fraction change in $Al_xGa_{1-x}N$. **c**, The changes of the per-mode contribution to ITC from four different classes of interface phonon modes with the Al fraction change in $Al_xGa_{1-x}N$.

# Methods

**Sample preparation.** The GaN/AlN heterostructures were fabricated by metal organic chemical vapor deposition (MOCVD). The AlN layer was grown by a typical two-step method using a home-made MOCVD system. Before the growth of AlN, the sapphire substrate was nitridated under the ammonia ($NH_3$) atmosphere with a flow of 4 slm at 1200 °C for 5 min. Subsequently, a thin AlN buffer layer was deposited at 900 °C. Then the AlN layer was grown at 1200 °C for 3 h by introducing $NH_3$ and trimethylaluminum (TMAl) with the flow of 500 sccm and 50 sccm, respectively. After that, a 100 nm-thick GaN layer was grown by MOCVD system (Veeco p125 turbo-disc vertical flow reactor) on the AlN layer at 1030 °C with an $NH_3$ flow of 4000 sccm and a trimethylgallium (TMGa) flow of 50 sccm. All growth processes adopted hydrogen ($H_2$) as the carrier gas.

The TEM specimens were thinned first by mechanical polishing and then by argon ion milling. The ion milling process was carried out using PIPS™ (Model 691, Gatan, Inc.) with the acceleration voltage of 3.5 kV until a hole was observed, followed by a low voltage milling with accelerating voltage of 0.3 kV to remove the surface amorphous layer and minimize the irradiation-damaged layers.

**STEM imaging.** HAADF and iDPC images were recorded at 300 kV using an aberration-corrected FEI Titan Themis G2. The convergence semi-angle for imaging was 30 mrad. The collection semi-angles snap was 4 to 21 mrad for the iDPC imaging and 39 to 200 mrad for the HAADF.

**EELS data acquisition.** The STEM-EELS were recorded using a Nion HERMES 200 microscope operating at 60 kV. For the off-axis experimental setup, the probe convergence semi-angle was 25 mrad and the collection semi-angle was 24.9 mrad. (Supplementary Fig. 1b). Beam deflection geometry ensures that scattered electrons with large momentum transfers are collected by the round spectrometer entrance aperture, which reduces the impact of phonon polaritons and exclude delocalized signal excited by dipole interaction[27,44]. Based on the scattering theory[28,29], the off-axis setup can also activate a larger variety of vibrational modes and the vibrational spectra will be preferably closer to phonon DOS. Moreover, by properly placing slot aperture at the diffraction plane, we can map high signal-noise-ratio phonon dispersion (Supplementary Fig. 1c) parallelly along high symmetry lines in Brillion zones (BZ)[39]. The uncertainty principle precludes the possibility to simultaneously obtain high spatial resolution and high momentum resolution. Thus, for enough momentum resolution, we employ a small convergence angle (~1.5 mrad) rather than the focused beam geometry, which scarifies a little spatial resolution but nanometer-scale is still achievable.

The thickness of the specimen in the regions of investigation was evaluated to be ~25 nm based on log-ratio methods[45]. The presented STEM-EELS dataset was acquired from a region of 1.2 × 8 nm containing the interface with 6×40 pixel mapping. The dwell time was 4 s / pixel with ~16 min for each dataset in total. Sample drift is usually less than 1 nm which is also corrected afterward by aligning the interface. The dispersion was 0.0005 eV/channel with typical energy resolution (full width at half maximum of the elastic line) 10 – 12 meV. Due to the collected region is larger than the first BZ of AlN and GaN, the obtained spectrum is closer to phonon DOS.

4D-EELS datasets were acquired with 1.5 mrad convergence semi-angle and a slot aperture with aspect ratio 16:1 placed along the ΓKMKΓ line. The corresponding momentum range was $0 < q < 10$ Å$^{-1}$ and the momentum resolution was ~ 0.2 Å$^{-1}$. The corresponding spatial resolution was ~ 4 nm. To avoid the central diffraction spot and enhance the signal-to-background ratio, the aperture was displaced along ΓKMKΓ direction by a reciprocal lattice vector. Typical dwell time was 15 s / pixel and ~ 32 min for each dataset in total with typical energy resolution 13 – 15 meV.

**EELS data analysis.** All acquired spectra were processed by custom-written Matlab code. For each dataset, EEL spectra were registered by their normalized cross correlation to correct beam energy drifts.

The extracted EEL spectra were also denoised by block-matching and 3D filtering (BM3D) algorithm[46]. The 4D-EELS datasets are individually denoised in two spatial dimensions for each energy and momentum channel. The noise level was estimated based on high-frequency elements in the Fourier domain. For 3D-EELS (off-axis results), the ZLP was removed by fitting the spectra to a Pearson function in two energy windows, one before and one after the energy loss region (approximately -10-12 meV and 130-160 meV). The peak position and intensity were extracted by least-squares fitting the background-subtracted spectra to a sum of multiple Lorentz peaks. The fitting was performed for each spectrum in the 3D dataset, and the standard deviation among rows were taken as the fitting uncertainty. For 4D-EELS datasets, we conducted a correction for the statistical factor following literature[47] in order to reduce the influence of elastic peak.

**Ab initio calculations.** The DFT calculations were performed within Quantum ESPRESSO[48] using "Perdew-Burke-Ernzerhof" exchange-correlation functional[49] and ultrasoft pseudopotential[50]. The kinetic energy cut-off was 44 Rydbergs (Ry) for wavefunctions and 395 Ry for charge density and potential. The AlN/GaN interface model contains 8 layers of GaN atoms connected to 8 layers of AlN (32 atoms in one hexagonal unit cell with cell parameters $a$ = 3.173 Å and $c$ = 41.048 Å). The structure was optimized the until the residual force was below $10^{-3}$ Ry per Bohr on every atom. The dynamical matrices and force constants were obtained using DFPT. The phonon dispersion and phonon DOS was calculated by interpolating the dynamical matrix on a 6 × 6 × 1 q-mesh. The AlN/Al$_x$Ga$_{1-x}$N was built with 7 × 8 in-plane expansion based on the above DFT-optimized atomic configuration (7 × 8 × 16 unit cell compare with AlN, 3584 atoms in total) and randomly replaced Ga atoms by Al atoms with ratio x in each atomic plane. The dynamical matrix was built under mass approximation, i.e. the interatomic force constants are assumed as invariant to composition ratio x. Then the dynamical matrix is diagonalized to get phonon frequencies and eigenvectors.

**Stokes cross-section of the scattering electrons.** The method calculated differential cross-section in the case of a fast electron interacting with phonons is described in a previous work[32].

**MD simulations.** The equilibrium MD simulations were performed using the LAMMPS package with the Stillinger-Weber (SW) potential [51,52]. The modal contribution to ITC was calculated within the framework of ICMA[18]. The modal contribution to ITC $G_n$ is calculated by the Green-Kubo formula

$$G_n = \sum_{n'} \frac{1}{A k_\mathrm{B} T^2} \int_0^\infty \langle Q_n(t) Q_{n'}(0) \rangle \, \mathrm{d}t$$

where $A$ is the interface area, $k_\mathrm{B}$ denotes the Boltzmann constant, $T$ is the temperature of the system, $Q_n(t)$ is the mode-decomposed heat current at time $t$ and angle brackets denote temporal correlation function averaged with different time origins.

The modal heat flux was calculated from modified LAMMPS[53]. The DFPT-calculated phonon eigenvectors were used as the basis vectors in modal decomposition. The mode-mode heat flux correlation was calculated by custom Matlab code. The result was averaged from 25 independent simulations for AlN/GaN, while for Al$_x$Ga$_{1-x}$N system the result was averaged from 5 randomly replaced atom configuration with 5 independent run each configuration. In each simulation, initial structure was firstly relaxed under temperature control at 300 K using velocity-rescaling. The heat flux was then calculated while running the simulation in the microcanonical ensemble for 350 ps with a time step of 1 fs. The modes were further classified to one of extended mode, partially extended mode, isolated mode or localized mode, by their amplitude distribution over space, similar to ref 23. The criteria for classification in AlN/B (B=GaN, Al$_x$Ga$_{1-x}$N) are as follows:

$$P_{i,tot} = \sum_{j \in entire\ system} |e_{ij}|$$

$$P_{i,int} = \sum_{j \in interface\ region} |e_{ij}|$$

$$P_{i,AlN} = \sum_{j \in AlN} |e_{ij}|$$

$$P_{i,B} = \sum_{j \in B} |e_{ij}|$$

Where $e_{ij}$ represents the vibration amplitude of atom $j$ and eigen mode $i$. The interface region contains the atoms within three atomic layers (~ 0.8 nm) from the interface plane. And the four types of interface modes are defined as

1) extended mode: $0.1 < P_{i,int}/P_{i,tot} < 0.28$, $0.1 < P_{i,AlN}/P_{i,B} \leq 10$;

2) partially extended mode: $0.1 < P_{i,int}/P_{i,tot} < 0.28$, $P_{i,AlN}/P_{i,B} \leq 0.1$ or $P_{i,AlN}/P_{i,B} > 10$;

3) isolated mode: $P_{i,int}/P_{i,tot} < 0.1$;

4) localized mode: $P_{i,int}/P_{i,tot} > 0.28$;

The respective contribution to ITC can thus be calculated by summing ITC over corresponding modes.